\documentclass[useAMS,usenatbib,fleqn]{mn2e}
\usepackage{amsmath,amssymb,amsfonts,latexsym}
\usepackage[dvips]{graphicx}
\usepackage{longtable}
\usepackage{fixltx2e}
\usepackage{natbib}
\usepackage{hyperref}

\newcommand{\be}{\begin{equation}}
\newcommand{\ee}{\end{equation}}

\def\be{\begin{equation}} 
\def\ee{\end{equation}} 
\def\beq{\begin{eqnarray}} 
\def\eeq{\end{eqnarray}}

\def\04a{{2004 a}}
\def\04b{{2004 b}}

\title[A proposed mechanism for radio emission state changes]{Rapid 
Modification of Neutron Star Surface Magnetic Field: A proposed mechanism for 
explaining Radio Emission State Changes in Pulsars}

\author[U. Geppert et. al.] {U. Geppert$^{1}$\thanks{E-mail:Ulrich.Geppert@dlr.de}, R. Basu$^{2}$, D. Mitra$^{3}$, G. I. Melikidze$^{1,4}$, M. Szkudlarek$^{1}$\\
$^{1}$ Janusz Gil Insitute of Astronomy, University of Zielona G\'ora, ul Szafrana 2, 65-516 Zielana G\'ora, Poland \\
$^{2}$ Inter-university centre for Astronomy and Astrophysics, Pune-411007, India\\
$^{3}$ National Centre for Radio Astrophysics, Tata Institute of Fundamental Research, Post Bag 3, Ganeshkind,Pune-411007,India\\
$^{4}$ Evgeni Kharadze Georgian National Astrophysical Observatory, 0301, Abastumani, Georgia }

\begin{document}

\date{}

\maketitle

\label{firstpage}

\begin{abstract}
The radio emission in many pulsars show sudden changes, usually within a 
period, that cannot be related to the steady state processes within the inner
acceleration region (IAR) above the polar cap. These changes are often 
quasi-periodic in nature, where regular transitions between two or more stable 
emission states are seen. The durations of these states show a wide variety 
ranging from several seconds to hours at a time. There are strong, small scale 
magnetic field structures and huge temperature gradients present at the polar 
cap surface. We have considered several processes that can cause temporal 
modifications of the local magnetic field structure and strength at the surface
of the polar cap. Using different magnetic field strengths and scales, and also 
assuming realistic scales of the temperature gradients, the evolutionary
timescales of different phenomena affecting the surface magnetic field was 
estimated. We find that the Hall drift results in faster changes in comparison 
to both Ohmic decay and thermoelectric effects. A mechanism based on the 
Partially Screened Gap (PSG) model of the IAR has been proposed, where the Hall
and thermoelectric oscillations perturb the polar cap magnetic field to alter 
the sparking process in the PSG. This is likely to affect the observed radio 
emission resulting in the observed state changes.
\end{abstract}


\section{Introduction}
A rapidly rotating, highly magnetized, neutron star is a unipolar inductor. The
region around the star, the magnetosphere, is initially charge starved and can 
develop extremely high corotation electric fields. In this environment a 
copious amount of charges from the neutron star as well as those created due to
magnetic pair production screens the electric field and form a force-free 
corotating magnetosphere. At the light cylinder, the magnetic field lines are 
separated into closed and open field line regions. The charges in the closed 
field line region corotate with the star, while a relativistic outflow 
consisting of dense pair plasma is established along the open field line 
region. The charges flow out of the neutron star in the form of a pulsar wind, 
and the magnetosphere requires a continuous supply of charges in order to 
maintain a steady outflow. This supply is generated by pair creation in the 
magnetosphere which ensures that within a few tens of microseconds the charges 
are replenished. The coherent radio emission arises due to growth of plasma 
instabilities in this relativistic outflow in the inner magnetosphere less than
10\% of the light cylinder radius \citep{M17}. As a result the observer 
samples the radio emission as a series of narrow single pulses which occupy 
less that 10\% of the pulsar period and each single pulse is highly variable. 
On the other hand several studies have demonstrated the pulsar profile, formed 
after averaging several thousand single pulses, to be highly stable 
\citep{HMT75,RR95}. This signifies the extreme stability of the averaging 
process over timescales of several minutes to hours. 

\begin{figure}
\centering
\includegraphics[scale=0.7,angle=0.]{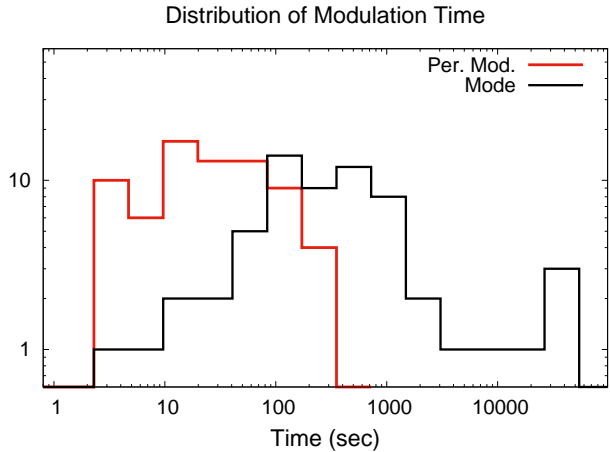}
\caption{The figure shows the distribution of timescales associated with
periodic and quasi-periodic modulations of radio emission from pulsars. The
x-axis represents the characteristic times of modulation while the y-axis
shows the number of pulsars within each time bin. The periodic modulations
(red histogram) consist of periodic nulling and periodic amplitude modulation
and are usually seen over a narrow range with periodicities between 10 and 100
seconds. The quasi-periodic variations (black histogram) include emission mode
changing as well as the remaining nulling pulsars, and shows a wide range
between few tens of seconds to several hours at a time.}
\label{fig1}
\end{figure}

Long term monitoring have revealed that in some pulsars the average profile can
vary over several different timescales, signifying a change in the emission 
state \citep[see e.g.][]{LHKSS10}. In most cases such variations are seen in 
normal long period pulsars, typically having periods longer than 30 
milliseconds, although recently similar behaviour has also been reported in 
young millisecond pulsars \citep{MKM18}. In this paper we will restrict our 
discussion to normal long period pulsars, and for easy identification we group 
these variations into Short-Timescale (ST) category, when the timescales of 
variations last from a single rotation period to a few hours, and 
Long-Timescale (LT) category, where the variations are observed over months to 
years \citep{KLOJL06,CRCJD12}. The LT category include intermittent pulsars, 
rotating radio transients and normal pulsars
 \citep[see, e.g.,][]{
2009ASSL..357...67L,2017ApJ...834...72L,2011BASI...39..333K}. 
The ST category
consist of three different phenomena, namely, mode changing, nulling and 
periodic modulations. Mode changing was first reported in the pulsar B1237+25 
by \citet{B70b}, where the emission switched between two separate states with 
different but stable profile shapes. The switching between the modes are very 
rapid and happens within a rotation period, but each mode can last for 
timescales ranging from minutes to hours. No clear periodic or quasi-periodic 
behaviour could be associated with mode transitions, although several pulsars 
have been identified recently, like J1752+2359 \citep{LWF04}, which exhibit 
quasi-periodic bursting states. During nulling the emission entirely ceases 
within a period which lasts over timescales from a single period to several 
minutes. There are also instances where the pulsar nulls for longer durations 
lasting hours at a time \citep{KHS14,NJMK18}. The periodic modulations are 
further subdivided into periodic amplitude modulations and periodic nulling. In
some cases the nulls are found to be periodic and repeats at intervals of 
several tens to a few hundred periods \citep{HR07,HR09,BMM17,BLK20}. The 
periodic amplitude modulation is another similar behaviour, where on short time
scales of tens to hundreds of periods the pulsar oscillates quasi-periodically 
between two distinct emission states usually of different intensity levels 
\citep{BMM16,MR17,YMW19,YMW20}. 

A detailed study of periodic modulation in the pulsar population was undertaken
in \cite{BMM20}, that found periodic amplitude modulation to be present in 18 
pulsars, while periodic nulling is seen in 29 pulsars. In addition there are 
another 24 pulsars where periodic modulation is also present, whose nature 
could not be resolved due to weaker detection sensitivity of the single pulses.
The distribution of the modulation periodicity of this population is shown in 
Fig. \ref{fig1} (red histogram), where the time has been estimated in seconds. 
The distribution is restricted to a relatively narrow window between 10-100 
seconds. There are around 30-40 pulsars where presence of mode changing have 
been detected. Table \ref{tabmode} in the appendix presents the details of the 
timescales associated with the known mode changing pulsars in the literature. 
The detection of nulling can be confused in weaker pulsars where low intensity 
single pulses are below detection limit. The presence of unambiguous nulling 
has been seen in around 80-100 pulsars \citep{WMJ07,GJK12,BMM20}. Although no 
underlying periodicity is associated with mode changing and remaining nulling 
cases, it is still possible to estimate the average durations of the different 
states. We were able to determine the average timescales associated with 
nulling and mode changing in around 60 pulsars whose distribution is shown in 
Fig. \ref{fig1} (black histogram). Unlike the periodic modulations, these 
timescales show a much wider variation between few seconds to several hours.

The radio emission only constitute a tiny fraction of the total rotational 
energy of the neutron star, and hence it is important to understand whether 
during the state changes there is any detectable change in the slowdown rate of
the star. For LT category, it has been shown that the profile changes are 
accompanied by changes in the spin down rate of the star \citep{KLOJL06}. This
opens up the possibility that changes in the global magnetospheric 
configuration can influence these state changes (\citealt{2010MNRAS.408L..41T,2014MNRAS.437..262M}). 
However, in this work we focus
on the ST phenomena, where no clear change in the global magnetosphere has been
observed during the state changes. Two mode changing pulsars, B0943+10 and 
B0823+26, have shown synchronous variations in the X-ray emission during radio
mode changes \citep{H13,HKB18}). Although the X-ray emission carries a higher 
fraction of the rotational energy compared to the radio emission, during mode 
transitions, the X-ray emission changes by only a small fraction, around 
15\%. This is not sufficient to constitute a change in the slowdown energy of
the star \citep{2017JApA...38...54M}). It has also been shown in several cases 
that in the different modes the radio emission arises from similar locations 
within the magnetosphere \citep{BMR19,RBMM21}. This further suggests that there 
is no change in the magnetic field structure or the open field line geometry in
the emission region during mode changing. The ST phenomena usually repeat at 
irregular intervals which also rules out highly periodic processes like 
precession and/or nutation as possible basis for such changes. Thus it is 
reasonable to suggest that the ST processes are governed by local changes in 
the magnetosphere that can effect a change in the radio emission.

Our aim in this work is to highlight 
a physical mechanism that can explain the ST phenomena in pulsars. 
Here we consider pulsar radio emission models such as \cite{RS75,AM98,GMG03,PTS20}, where 
the generation of coherent radio emission is facilitated by nonstationary 
pair plasma discharges above the pulsar polar cap.  
This can be best 
done in the framework of the partially screened gap \citep[PSG,][]{GMG03} model, which describes the physics
 immediately above the polar cap. We suggest that changes in the surface magnetic field 
on the polar cap over short timescales can bring about changes in the radio 
emission. We briefly describe below the physical conditions on the polar cap 
that that can sustain such a mechanism.

\subsection{The properties of polar cap influencing coherent radio emission}
A widely accepted model for the origin of pair plasma responsible for the 
coherent radio emission has been proposed by \citet{RS75}, which suggests the 
existence of an inner acceleration region (IAR) above the polar cap. A system 
of sparking discharges driven by magnetic pair production is setup in the IAR
that supplies a non-stationary flow of relativistic outflowing plasma along 
the open field line region. Several observational studies suggest that the IAR 
is dominated by strongly non-dipolar surface magnetic field \citep[e.g.][]{
2019MNRAS.489.4589A} and as a consequence is not a pure vacuum gap as was 
initially assumed by \cite{RS75}, but form a PSG
\citep{GMG03} due to supply of thermally regulated ions from the 
stellar surface. The strong non-dipolar nature causes the field lines to become
highly curved, thereby enhancing the generation and acceleration of the dense 
pair plasma in the PSG. The large electric fields in the gap separates the pair
plasma in opposite directions. The electrons smash against the polar cap 
surface and heats it up which causes the iron ions to escape the surface into
the gap. The positrons and ions are accelerated outwards by the electric field 
to relativistic speeds. They produce additional pair cascades which result in 
a secondary pair plasma flowing outwards along the open field lines. The radio 
emission is generated in this outflowing plasma due to growth of plasma 
instabilities \citep{AM98,MGP00,LMM18,RMM20}.

\subsection{The temperature variations on the Neutron star surface}
The heated polar cap surface is expected to produce thermal X-ray emission 
which can in turn be used to find its area and temperature. Studies of 
time-aligned radio and thermal X-ray emission reveal that the polar cap region 
is dominated by strong small scale field and hence is smaller than what is 
expected from a purely dipolar configuration \citep{GGM13,SGZHMGMX17}. Further 
the polar cap temperatures have been measured to exceed $3\times 10^6$ K. It 
has also been shown that a large temperature difference exist between the polar
cap and the rest of the neutron star surface. This results in enormous 
temperature gradients both in radial and meridional directions between the 
polar cap envelope and the surrounding regions, which are restricted to within 
a very shallow layer beneath the surface. The kinetic energy of the backflowing 
charges is released as heat within a few radiation lengths and is immediately 
re-radiated, without practically any thermal energy being transported deeper 
into the crust \citep{CR80,SG20}. The lifetime of radio pulsars are typically 
around $10^6$ to $10^7$ years, during which the polar cap area remains 
significantly hotter than the remaining surface. Since the magnetic field 
penetrating the polar cap surface is dominated by strong radial components, 
very effective thermal insulation is present at the rim of the polar cap. The 
remaining surface cools in accordance with quite well understood processes 
(URCA or DURCA, photons), thereby increasing the meridional temperature 
gradient with time. After $\sim 10^6$ years the remaining surface has 
temperatures around few times $10^5$ K \citep{PGW06,VRPPAM13}.

\subsection{Magnetic field variations at small spatial scales on the Neutron 
star surface}
The magnetic field at the polar cap is maintained by small scale currents 
circulating beneath its surface. These currents are generated and maintained by
the continuous transfer of magnetic flux from a large scale toroidal field 
reservoir into small scale poloidal structures via the Hall drift 
\citep{RBPAL07,GV14}. The formation of such magnetic spots has been recently 
demonstrated in 3-dimensional simulations by \cite{GH18}. The magnetic spots 
are confined to small spatial regions, and extreme thermal insulation of the 
polar cap by the strong local poloidal field ensure that the magnetic field 
variations in these small regions are independent of the global magneto-thermal
evolution of the neutron star. 

In this paper we suggest that the Hall drift and thermoelectrically driven 
magnetic field oscillations at the polar cap surface is a possible explanation
for the observed ST phenomena in pulsars. Estimates of the timescales 
associated with these processes have been carried out which match the 
observations. We also present a mechanism where these changes in the surface 
field affect the sparking process in the polar cap thereby inducing changes in 
the pulsar radio emission in a quasi-periodic manner. In section~\ref{sec2} we
give the framework for the variations in the magnetic field and estimate the 
timescales involved. In section~\ref{sec3} and ~\ref{sec4} we propose a model 
for the ST phenomena and present our conclusions.

\section{Timescale Analysis of magnetic field variation over polar cap}
\label{sec2}

\subsection{Basic equations}
The surface of the polar cap is made up of highly conducting iron nuclei that 
are either in a liquified or crystallized state. A striking feature of this 
region is the presence of extremely large temperature gradients; especially in 
radial direction as well as the meridional direction. The magnetic field 
evolution is governed by the induction equation :
\beq
\frac{\partial\vec{ B}}{\partial t} =-c\vec{\nabla}\times\vec{E}\;,
\label{eq:IndEqGeneral}
\eeq
\noindent 
where the electric field $\vec{E}$ is given by Ohm's law and is a sum of three 
terms:
\beq
\vec{E}=\frac{c}{4\pi\hat{\sigma}}\vec{\nabla}\times\vec{B}-\frac{\vec{v}_{adv}}{c}\times\vec{B}-\vec{E}^{ext}\;.
\label{eq:ElecField}
\eeq 
\noindent 
The first term contains the tensor of the electric conductivity $\hat{\sigma}$.
The components of $\hat{\sigma}$ that are perpendicular to the direction of the
magnetic field are responsible for the Hall drift of the magnetic field. The 
first term also contains the effect of the usual Ohmic diffusion. The second 
term in Eq.~\ref{eq:ElecField} is the advection term where $\vec{v}_{adv}$ 
is the advection velocity and this term is relevant only when the matter at the polar 
cap is in a liquified state, which is possible when the temperatures are 
extremely high. Such a situation is unlikely to be realized in radio pulsars 
considered here (see Sect.~\ref{sec:3.1}). The nature of the external electric 
field $\vec{E}^{ext}$, which constitute the third term, is determined by the 
presence of huge temperature gradients. These gradients are especially strong 
in the radial direction where they can exceed $\sim10^8$ Kcm$^{-1}$. Across the
rim of the polar cap the meridional temperature gradients of $\sim 10^4$ 
Kcm$^{-1}$ may drive a thermoelectric instability which causes the generation 
of an azimuthal magnetic field along that rim \citep{BAH83, ULY86, GW95}. 
In the context of emission mode changing the thermal drift of the magnetic 
field by these temperature gradients are of special interest. It may cause an 
oscillating behaviour of the local magnetic field structures at the real polar 
cap surface.

If at the polar cap $\vec{v}_{adv}=0$, the two remaining electric field 
constituents in Eq.~\ref{eq:ElecField} have the following form:
\beq
\frac{c^2}{4\pi\hat{\sigma}}\vec{\nabla}\times\vec{B}=\eta_B\vec{\nabla}\times\vec{ B}+\alpha_H(\vec{\nabla}\times\vec{ B})\times \vec{B}
\nonumber\\
\eta_B=\frac{c^2}{4\pi\sigma}\;,\;\; \alpha_H=\eta_B\frac{\omega_B\tau}{B}\;,
\label{eq:OhmHall}
\eeq
\noindent 
where $\sigma$ is the electric conductivity tensor component parallel to 
$\vec{B}$, $\omega_B\tau$ the magnetization parameter, and $B$ the magnitude of
the magnetic field vector $\vec{B}$. The thermoelectric field is given as:
\beq
\vec{E}^{ext}=\vec{E}_T&=&-Q(3+\xi)\vec{\nabla}T
\nonumber\\
&+&\frac{Q\xi\omega_B\tau}{1+\omega_B^2\tau^2}\left[\vec{\nabla}T+\omega_B\tau(\vec{b}\times\vec{\nabla}T)\right]\times \vec{b}\;.
\label{eq:TE}
\eeq
\noindent 
In these equations $\vec{b}$ is the unit vector along $\vec{B}$ and 
$\omega_B\tau$ is the magnetization parameter, indicating the strength of 
magnetization of the electron plasma. The magnetic diffusivity $\eta_B$ depends
on $\sigma$. The thermopower $Q$, the coefficient $\xi$, as well as the 
derivation of Eq.~\ref{eq:TE} are explained in detail in the 
Appendix~\ref{sec:8.4}. The first term in the r.h.s. of Eq.~\ref{eq:TE} 
describes the thermoelectric battery. If the coefficient $Q(3+\xi)$ depends on 
the radial coordinate $r$ and a meridional temperature gradient $dT/rd\theta$ 
is present, this effect may amplify an azimuthal seed field. For the study of 
local magnetic field oscillations the second term is of interest. It can be 
interpreted, similar to the Hall drift, as the thermal drift of the magnetic 
field with the thermal drift velocity, $\vec{v}_{TD}/c\times \vec{B}$, where 
\be
\vec{v}_{TD}=\frac{Q\xi e\tau}{m_{\ast}(1+\omega_B^2\tau^2)}\left[\vec{\nabla}T+\omega_B\tau(\vec{b}\times\vec{\nabla}T)\right]\;.
\label{eq:v_TD}
\ee
\noindent 
In the limit $\omega_B\tau\gg1$, which is expected to be valid in the polar cap
region of radio pulsars, $\vec{E}_T$ simplifies to
\be
\vec{E}_T=-Q(3+\xi)\vec{\nabla}T + Q\xi(\vec{b}\times\vec{\nabla}T)\times \vec{b}.
\label{eq:E_T}
\ee
\noindent 
The $\mathrm{curl}$ of Eqs.~\ref{eq:OhmHall} and~\ref{eq:E_T} leave us with the
induction equations  govern the magnetic field evolution by Ohmic diffusion, 
the Hall drift, and thermoelectric effects.

The magneto-thermal evolution at the polar cap surface is a complex non-linear 
process whose detailed interpretation can be obtained only from numerical 
modelling. However, in order to get a basic idea of the processes from 
analytical results we will use the following simplifications:
\begin{enumerate}
 \item Axial symmetry is assumed. This greatly simplifies the use of poloidal 
- toroidal representation of the magnetic field: $\vec{B}=\vec{B}_p + 
\vec{B}_t$. Under this assumption the poloidal and toroidal components of 
$\vec{B}$ have the following properties:
\begin{enumerate}
\item $\vec{\nabla}\times\vec{B}_p$ is toroidal, $\vec{\nabla}\times\vec{B}_t$ is poloidal,
\item $\vec{B}_p \cdot \vec{B}_t=0$,
\item $\vec{B}_p \times \vec{B}_t$ is poloidal,
\item $\vec{B}_p \times \vec{B}_p=0$,
\item $\vec{B}_t \times \vec{B}_t=0$.
\end{enumerate}
\item  The magnetic field evolution at the polar cap surface take place within 
a very thin layer. Therefore, we will consider all transport coefficients as 
independent of coordinates.
\end{enumerate}
\noindent Without the above simplifications one would require much less illustrative 
dispersion relations to describe the physical behaviour. But it is unclear 
whether any essential physics is neglected by using these assumptions.

\noindent As shown below, the advective term in Eq.~\ref{eq:ElecField} can be neglected in our analysis.
The remaining constituents of the elcetric field $\vec{E}$ in Eq. ~\ref{eq:IndEqGeneral},
given by Eq.  ~\ref{eq:OhmHall} and Eq. ~\ref{eq:TE} will cause different temporal variations of the
magnetic field $\vec{B}$. One would expect that the Hall induction equation, related to the
tensor of the electric conductivity describes a faster magnetic field variation than the 
thermoelectric induction equation related to $\vec{E}_T$. 
Under the above given simplifications, the Hall induction equation is given as:
\beq 
\left(\frac{\partial\vec{ B}}{\partial t}\right)_{Ohm,Hall} = -\eta_B\vec{\nabla}\times\left[\vec{\nabla}\times\vec{ B}\right]
+\alpha_H\vec{\nabla}\times\left[ (\vec{\nabla}\times\vec{ B})\times \vec{B}\right]\;,
\label{eq:HallIndEq}
\eeq
while the effect of temperature gradients on the magnetic field evolution is 
of the form:
\beq
\left(\frac{\partial \vec{ B}}{\partial t}\right)_{TE} =-\vec{\nabla}\times\left[\alpha_T\vec{\nabla}T-
\beta_T\left(\vec{ B}\times \vec{\nabla}T\right)\times \vec{ B}\right]\;,
\label{eq:TEIndEq}
\eeq
where $\alpha_T=cQ(3+\xi)$ and $\beta_T=cQ\xi/B^2$. The first term in the 
r.h.s. of Eq. ~\ref{eq:TEIndEq} describes effect of thermoelectric battery. It 
is independent of $\vec{B}$ and could contribute only to a toroidal seed field.
Since $\vec{\nabla}\times\left(\alpha_T\vec{\nabla}T\right)=\vec{\nabla}\alpha_T \times \vec{\nabla}T$, this term vanishes in our approximation of coordinate
independent transport coefficients. Thus, in this study we will calculate the 
dispersion relations of magnetic perturbations, whose evolution is determined 
by Eqs.~\ref{eq:HallIndEq} and~\ref{eq:TEIndEq}. But before that we will discus
the physical conditions at the polar cap.

\subsection{Conditions of aggregation at the polar cap surface}
\label{sec:3.1}
An important issue relevant to this work is the surface density at the hot 
polar cap, where the neutron star matter is condensed, either in liquified or 
solidified state. The most recent description of the state of aggregation in 
this region is provided in \cite{PC13}. The so called zero pressure density can
be considered at the surface which is bombarded with ultra-relativistic 
electrons. It increases with the local magnetic field strength, $\propto 
B_{12}^{6/5}$, and is estimated as:
\beq
\rho_s=561\zeta A Z^{-3/5}B_{12}^{6/5}\,\,, \zeta \approx 0.517+0.24 B_{12}^{1/5}\, ,
\label{eq:rho_s}
\eeq
where $B_{12}=B/10^{12}$G. In case of a surface made up of Fe ($A=56, Z=26$), 
$\rho_s\approx 5.1\times10^5, 1.25\times10^6, ...$ g~cm$^{-3}$ for local field
strengths $B_{12}= 50, 100, ...$, respectively, as expected at the non-dipolar 
polar cap surface. The Coulomb coupling parameter, $\Gamma_m(B)$, which depends
on the magnetic field, determines the state of surface matter, i.e., whether it 
is liquid or solid. If $\Gamma_m(B) > 175$ the matter is in a solidified form, 
while below this value it is a liquid. The parameter is estimated in 
\cite{PC13} as:
\beq
\Gamma_m(B)\approx \Gamma(0)\left[1+0.2\beta\right]^{-1}\, .
\label{eq:GammaB}
\eeq
Here $\beta$ is given by the ratio of ion cyclotron to ion plasma frequency, 
$\beta\approx 0.0094 B_{12}\rho_6^{-1/2}$ with $\rho_6=\rho/10^6$ gcm$^{-3}$. 
$\Gamma(0)$ is roughly the ratio of Coulomb and thermal energies,
\beq
\Gamma(0)=\frac{(Ze)^2}{a_ik_BT}\approx \frac{22.747}{ T_6}\rho_6Z^2A^{-1/3}\, ,
\label{Gamma0}
\eeq
with $a_i$ being the spacing between ions and $k_B$ is the Boltzmann constant. 
Inserting typical values for the surface temperature at the polar cap, 
$T_6\approx 3, 5$ \citep[see e.g.][]{SGZHMGMX17} and $\rho_6\approx 0.51, 
1.25$, one finds $\Gamma_m \approx 974, 1235$ if $T_6=3$ and $\Gamma_m \approx 
400, 1000$ if $T_6=5$ for $B _{12}= 50, 100$, respectively. In all cases 
$\Gamma_m(B)$ is significantly larger then $175$. Hence, the polar cap surface 
of radio pulsars is in a solidified state. For this reason an advection of the 
polar cap magnetic field is impossible and the corresponding term $\nabla\times
(\frac{\vec{v}_{adv}}{c}\times\vec{ B})$ in Eq.~\ref{eq:ElecField} can be 
safely neglected.

\subsection{Local magnetic field structure at the polar cap surface}
\label{sec:3.2}
An useful approximation for the non-dipolar magnetic field on the polar cap 
surface has been suggested by \cite{GMM02}, who consider the field to be a 
superposition of a star centered global dipole and a crust anchored surface 
dipole, $\vec{B}_{s} = \vec{B}_{d} + \vec{B}_{p}$. The magnetic field near the
polar cap is dominated by the surface dipole, while at large distances from the
star the global dipole takes precedence. In this configuration the magnetic 
field structure near the surface differs greatly from a purely dipolar one, 
but the field lines reconnects with the dipolar field at altitudes of about few
hundred meters. We have used the surface dipole to be characterized as 
$\vec{B}_{p}=B_{p0}\frac{R_{pc}^3}{r^3}\left(\mathrm{cos}\theta\hat{e}_r+\frac{1}{2}\mathrm{sin}\theta\hat{e}_{\theta}\right)$, where $R_{pc}$ is the dipolar 
polar cap radius. The global dipole has typical strength of a few times 
$10^{12}$ G at the polar cap surface. The local dipole is expected to be 
located at a depth of $\sim 5\times 10^3$ cm below the crust, which is 
approximately the size of the non-dipolar polar cap radius $R_{pc}$ \citep[see 
Table 1~in][]{G17}. Using $rd\theta \approx 5\times 10^3$ cm, the meridional 
angle directed from the pole of $\vec{B}_{p0}$ to the rim of the polar cap is 
$\theta \approx 45^{\circ}$.

\subsection{Hall oscillations}
We follow the method presented in \cite{SU97} to derive the dispersion relation
for coupled Ohmic decay and Hall drift. In the presence of strong magnetic 
fields the electric conductivity becomes a tensor. 
The decomposition of the magnetic field into poloidal and toroidal components 
transforms the induction Eq.~\ref{eq:HallIndEq} into coupled differential equations of the 
form:
\beq
\frac{\partial\vec{ B}_p}{\partial t}& =&-\eta_B\left[\vec{\nabla}\times(\vec{\nabla}\times\vec{B}_p)\right]
\nonumber\\
&-&\alpha_H\vec{\nabla}\times\left[(\vec{\nabla}\times\vec{B}_t) \times \vec{B}_p\right]\;,\;
\label{eq:HallIndEqPol}
\eeq
\beq
\frac{\partial\vec{ B}_t}{\partial t}& =&-\eta_B\left[\vec{\nabla}\times(\vec{\nabla}\times\vec{B}_t)\right]
\nonumber\\
& -&\alpha_H\vec{\nabla}\times\left[(\vec{\nabla}\times\vec{B}_p) \times \vec{B}_p+(\vec{\nabla}\times\vec{B}_t)\times\vec{B}_t\right]\;.
\label{eq:HallIndEqTor}
\eeq
We introduce small perturbations $\delta\vec{B}_p, \delta\vec{B}_t$, to the 
background magnetic field components $\vec{B}_{p0}, \vec{B}_{t0}$, similar to 
\cite{SU97}, such that $\vec{B}_p=\vec{B}_{p0}+\delta\vec{B}_p$ and 
$\vec{ B}_t= \delta\vec{B}_t $ ($\vec{B}_{t0}=0$ is assumed). The 
Hall-Induction equations up to first order in the perturbations are obtained as:
\beq
\frac{\partial \delta\vec{ B}_p}{\partial t} = -\eta_B\left[\vec{\nabla}\times(\vec{\nabla}\times\delta\vec{B}_p)\right] - \alpha_H\vec{\nabla}\times\left[(\vec{\nabla}\times\delta\vec{B}_t) \times \vec{B}_{p0}\right]\;, 
\label{eq:PertPol}
\eeq
\beq
\frac{\partial \delta\vec{ B}_t}{\partial t} = -\eta_B\left[\vec{\nabla}\times(\vec{\nabla}\times\delta\vec{B}_t)\right] - \alpha_H\vec{\nabla}\times\left[(\vec{\nabla}\times\delta\vec{B}_p) \times \vec{B}_{p0}\right]\;,
\label{eq:PertTor}
\eeq
The differential equations are transformed into algebraic ones by using Fourier 
transformation. The process is described in Appendix~\ref{sec:FT} for the Ohmic
decay terms in Eqs.~\ref{eq:PertPol} and \ref{eq:PertTor}. If we consider the 
Hall term for poloidal perturbation in Eq.\ref{eq:PertPol} which is of the form
\beq -\alpha_H\vec{\nabla}\times\left[(\vec{\nabla}\times\delta\vec{ B}_{t})\times\vec{B}_{p0}\right]\;,
\label{eq:HallPol}
\eeq
the corresponding expression in Eq.~\ref{eq:HallIndEqPol} is 
$\alpha_H (\vec{k}\vec{B}_{p0})\times\delta\vec{ B}_{t}$. Applying the same 
procedure to the Hall term of Eq.~\ref{eq:HallIndEqTor} we get 
$\alpha_H (\vec{k}\vec{B}_{p0})\times\delta\vec{ B}_{p}$. This leads to a set of 
two algebraic equations
\beq
\left(i\omega+\eta_B k^2\right)\delta\vec{ B}_{p}=-\alpha_H(\vec{k}\cdot\vec{B}_{p0})\vec{k}\times\delta\vec{ B}_{t}\;,\\
\left(i\omega+\eta_B k^2\right)\delta\vec{ B}_{t}=-\alpha_H(\vec{k}\cdot\vec{B}_{p0})\vec{k}\times\delta\vec{ B}_{p}\;.
\label{eq:CoupledHall}
\eeq
This system of equations can be easily solved by substituting 
\be
\delta\vec{ B}_{t}=-\alpha_H\frac{(\vec{k}\cdot\vec{B}_{p0})\vec{k}}{(i\omega+\eta_B k^2)}\times\delta\vec{ B}_{p}\,,
\label{eq:TorHall}
\ee
to find the dispersion relation for both the toroidal and poloidal perturbation as 
\be
\omega_{\pm}=i\eta_B k^2 \pm \alpha_H(\vec{k}\cdot\vec{B}_{p0})k=i\eta_B k^2\pm \omega_{Hall}\,.
\label{eq:DispHall}
\ee
The first term in the r.h.s. of Eq. ~\ref{eq:DispHall} describes the (perhaps 
slow) Ohmic decay of the Hall wave amplitudes, while the second term is the 
frequency of the sinusoidal Hall waves, sometimes called helicoidal or Whistler
waves. As pointed out in \cite{SU97}, this Hall waves may appear in strong 
magnetic fields, $\vec{k}\cdot\vec{B}_{p0}\gg 1$, with relatively small 
wavelengths, which is applicable for the polar cap of radio pulsars.

\subsection{Thermal drift oscillations}
In the limit $\omega_B\tau\gg1$ the thermal electric field, $\vec{E}_T$, 
simplifies to the form (see Appendix~\ref{sec:8.4}, Eq.~\ref{eq:E_T2} and 
\ref{eq:E_T_simpl})
\be
\vec{E}_T=-Q(3+\xi)\vec{\nabla}T + Q\xi(\vec{b}\times\vec{\nabla}T)\times \vec{b}.
\label{eq:E_T3}
\ee
The curl of $\vec{E}_T$ gives the `thermal' induction equation
\beq
\frac{\partial \vec{ B}}{\partial t} =-\vec{\nabla}\times\left[\alpha_T\vec{\nabla}T-
\beta_T\left(\vec{ B}\times \vec{\nabla}T\right)\times \vec{ B}\right]\;.
\label{eq:IndEqTherm}
\eeq
where $\alpha_T=cQ(3+\xi)$ and $\beta_T=cQ\xi/B^2$.  The first term in the 
r.h.s. of Eq. ~\ref{eq:IndEqTherm} is the thermoelectric battery term. It is 
independent of $\vec{B}$ and can only contribute to the toroidal background 
field $\vec{B}_{t0}$. Since $\vec{\nabla}\times\left(\alpha_T\vec{\nabla}T\right)=\vec{\nabla}\alpha_T \times \vec{\nabla}T$, this term vanishes in our 
approximation of coordinate independent transport coefficients, and only the 
remaining term in the r.h.s. of Eq. ~\ref{eq:IndEqTherm} describes the thermal 
drift of magnetic field. The drift velocity $\vec{v}_{TD}=\beta_T\left(\vec{ B}\times \vec{\nabla}T\right)$ depends on the magnetic field and can affect 
the field evolution in the presence of strong temperature gradients, which are
typically present both in radial and in meridional direction of the polar cap 
surface.

Assuming $\beta_T$ to be independent of coordinates, and applying the 
decomposition of the magnetic field into poloidal and toroidal components we 
have
\beq
\frac{\partial \vec{ B}}{\partial t} &=&-\beta_T\vec{\nabla}\times\left[(\vec{B}_p\times \vec{\nabla}T)\times\vec{B}_p\right]
\nonumber\\
&-&\beta_T\vec{\nabla}\times\left[(\vec{B}_t\times \vec{\nabla}T)\times\vec{B}_p\right]
\nonumber\\
&-&\beta_T\vec{\nabla}\times\left[(\vec{B}_t\times \vec{\nabla}T)\times\vec{B}_t\right]\;.
\label{eq:IndEqTherm1}
\eeq
The first and third terms in Eq.~\ref{eq:IndEqTherm1} are toroidal fields while
the second term is poloidal in nature. Thus, the thermal drift $\sim 
\vec{\nabla}T$ acts directly on both $\vec{B}_t$ and $\vec{B}_p$. Following the 
same procedure as above for the dispersion relation of Hall waves, the 
induction equations till first order in $\delta\vec{ B}_{p,t}$ are given as:
\beq
\frac{\partial \delta\vec{ B}_p}{\partial t}=-\beta_T\vec{\nabla}\times\left[(\delta\vec{ B}_t\times T')\times\vec{B}_{p0}\right]\;,
\label{eq:PolTherm}
\eeq
\beq
\frac{\partial \delta\vec{ B}_t}{\partial t}=-\beta_T\vec{\nabla}\times\left[(\delta\vec{ B}_p\times T')\times\vec{B}_{p0}+(\vec{B}_{p0}\times T')\times \delta\vec{ B}_p\right]\;,
\label{eq:TorTherm}
\eeq
where $T'$ denotes $\vec{\nabla}T$, a quantity which is assumed to be 
unperturbed over timescales of magnetic field variations. We first consider the
evolution of the poloidal component of the magnetic field perturbation because it may
affect the emission regime above the polar cap surface:
\beq
\frac{\partial \delta\vec{ B}_p}{\partial t}&=&\beta_T\vec{\nabla}\times\left[\vec{B}_{p0}\times (\delta\vec{ B}_t\times T')\right]
\nonumber\\
&=&\beta_T\vec{\nabla}\times\left[\delta\vec{ B}_t(\vec{B}_{p0}\cdot T')-T'(\vec{B}_{p0}\cdot\delta\vec{ B}_t)\right]\;.
\label{eq:PolTherm1} 
\eeq
The last term $\vec{B}_{p0}\cdot\delta\vec{ B}_t$ vanishes because in axial 
symmetry the scalar product of a poloidal and a toroidal vector is zero. The 
curl of the remaining equation gives 
\beq
\frac{\partial \delta\vec{ B}_p}{\partial t}=\beta_T\left[(\vec{B}_{p0}\cdot T') \vec{\nabla}\times\delta\vec{ B}_t+\vec{\nabla}(\vec{B}_{p0}\cdot T')\times \delta\vec{ B}_t\right]\;.
\label{eq:PolTherm2}
\eeq
We have used the abbreviation $m=(\vec{B}_{p0}\cdot T')$. Fourier 
transformations on the above equation gives us the relation
\beq
i\omega \delta\vec{ B}_p=\beta_T(m'-im\vec{k})\times \delta\vec{ B}_t\;.
\label{eq:PolTherm3}
\eeq
In order to solve this equation we couple the poloidal perturbation caused by $\vec{v}_{TD}$ to
the toroidal perturbation caused by the Hall drift, i.e. we couple the thermal drift oscillations to
the rapidly varying Hall drift oscillations. 
If $\delta\vec{ B}_t$ is replaced by the expression in Eq.~\ref{eq:TorHall}, 
\beq
\delta\vec{ B}_t=-\frac{n}{(i\omega+\eta_B k^2)}\vec{k}\times\delta\vec{ B}_p\;
\label{eq:TorHall1}
\eeq
where $n=\alpha_H (\vec{k}\cdot \vec{B}_{p0})$, into Eq. ~\ref{eq:PolTherm3} and 
rearranging the vector product we obtain
\beq
i\omega(i\omega+\eta_B k^2)\delta\vec{ B}_p&=&n\beta_T\left[(m'-im\vec{k})\times \delta\vec{ B}_p\right]\times\vec{k}
\nonumber\\
&=&n\beta_T\left[(m'-im\vec{k})\cdot\vec{k}\right]\delta\vec{ B}_p\;,
\label{eq:PolTherm4}
\eeq
as in plane waves $\delta\vec{ B}_p\cdot \vec{k}=0$ and we assume  $m' \parallel \vec{k}$ . Solving the above 
quadratic equation gives
\beq
\omega_{\pm}=i\frac{\eta_B k^2}{2}\pm\sqrt{-\frac{\eta_B^2k^4}{4}-n\beta_Tm'\cdot\vec{k}+in\beta_Tmk^2}\;.
\label{eq:DispTD}
\eeq
The real part of $\omega_{\pm}$ describes the frequency of the thermal drift 
oscillations of the poloidal magnetic field component. The real part of 
$\omega_{\pm}$ has the form (see Appendix~\ref{sec:compl_number}) 
\beq
\omega_+\approx \frac{\eta_B k^2}{2}\sqrt{\frac{4n\beta_T(m'\cdot\vec{k})}{\eta_B^2 k^4}+1}\;.
\label{eq:omega_plus}
\eeq

We have termed the scalar product of the background poloidal magnetic field and 
the temperature gradient, $m=(\vec{B}_{p0}\cdot T')$, as `magnetic temperature 
gradient'. This quantity is always negative at the polar cap surface, because 
the temperature within the polar cap decreases both with increasing radial and 
meridional coordinates, $dr$ and $rd\theta$. In the following analysis we use
spherical coordinates related to the center of the local dipole (see 
Sec.~\ref{sec:3.2}). Assuming $m' \parallel \vec{k}$ we have to first study 
$m'=\vec{\nabla}(\vec{\nabla} T\cdot \vec{B}_{p0})$. Clearly, $m'$ contains 
first derivatives of the magnetic field components as well as the first and 
second derivatives of temperature along the radial and meridional coordinates.
We evaluate $m'$ in spherical coordinates using axial symmetry, and for 
simplicity replace $\vec{B}_{p0}$ with $\vec{B}$
\beq
\vec{\nabla} T\cdot \vec{B}&=&\left(\frac{\partial T}{\partial r}\hat{e}_r+\frac{1}{r}\frac{\partial T}{\partial \theta}\hat{e}_{\theta}\right)\cdot(B_r\hat{e}_r+B_{\theta}\hat{e}_{\theta})
\nonumber\\
&=& B_r\frac{\partial T}{\partial r}+\frac{\frac{\partial T}{\partial \theta}}{r}\frac{\partial T}{\partial \theta}\;.
\eeq
From $\vec{\nabla}(\vec{\nabla} T\cdot\vec{B})=(\vec{\nabla} T\cdot\vec{\nabla})\vec{B}+(\vec{B}\cdot\vec{\nabla})\vec{\nabla} T+\vec{\nabla} T\times(\vec{\nabla}\times\vec{B})\;,  (\vec{B}\times (\vec{\nabla}\times(\vec{\nabla} T)))=0$ 
because $\vec{\nabla}\times(\vec{\nabla} T)=0$, we find for the $r$-component 
of $m'=\vec{\nabla}(\vec{\nabla} T\cdot \vec{B})$:
\beq
\frac{1}{r^2}\frac{\partial T}{\partial \theta}B_{\theta}+\frac{1}{r}\frac{\partial T}{\partial \theta}\frac{\partial B_{\theta}}{\partial r} + \frac{\partial T}{\partial r}\frac{\partial B_{r}}{\partial r}
\nonumber\\
 + B_r\frac{\partial^2 T}{\partial r^2} +\frac{B_{\theta}}{r}\frac{\partial^2 T}{\partial r \partial \theta}\;,
\label{eq:mdash-r}
\eeq
and for the $\theta$-component of $m'=\vec{\nabla}(\vec{\nabla} T\cdot \vec{B})$:
\beq
-\frac{1}{r}\frac{\partial T}{\partial r}B_{\theta} + \frac{1}{r}\frac{\partial T}{\partial r}\frac{\partial B_{r}}{\partial \theta} + \frac{1}{r^2}\frac{\partial T}{\partial \theta}\frac{\partial B_{\theta}}{\partial \theta}
\nonumber\\
+\frac{B_r}{r}\frac{\partial^2 T}{\partial r \partial \theta} - \frac{B_r}{r^2}\frac{\partial T}{\partial \theta} +\frac{B_{\theta}}{r^2}\frac{\partial^2 T}{\partial \theta^2}\;.
\label{eq:mdash-theta}
\eeq
Using specifications of the local dipole field at the surface as $r=R_{pc}=5\times 10^3$ cm, $\vec{B}_{p0}=B_{p0}(\mathrm{cos}{\theta}\hat{e}_r+\frac{1}{2}\mathrm{sin}{\theta}\hat{e}_{\theta})$, the derivatives of the magnetic field 
appearing in Eqs.~\ref{eq:mdash-r} and \ref{eq:mdash-theta} are 
\beq
\frac{\partial B_{r}}{\partial r}&=&-3B_{p0}\frac{R_{pc}^3}{r^4}(\mathrm{cos}{\theta})\;,
\nonumber\\
\frac{\partial B_{\theta}}{\partial r}&=&-3B_{p0}\frac{R_{pc}^3}{2r^4}(\mathrm{sin}{\theta})\;,
\nonumber\\
\frac{\partial B_{r}}{\partial \theta}&=&-B_{p0}\frac{R_{pc}^3}{r^3}(\mathrm{sin}{\theta})\;,
\nonumber\\
\frac{\partial B_{\theta}}{\partial \theta}&=&B_{p0}\frac{R_{pc}^3}{2r^3}(\mathrm{cos}{\theta})\;.
\eeq
Note that at the rim of the polar cap, $R_{pc}$, $\theta \sim 45^{\circ}$. 
Clearly, at $r=R_{pc}$ and in the vicinity of the magnetic pole 
$\frac{\partial B_{r}}{\partial \theta}$ is the largest partial derivative of 
the background magnetic field. For the primary contributions to $m'$ we find
from the first two terms of the $\theta$ component of $m'$ (see 
Eq.~\ref{eq:mdash-theta})
\beq
m'\approx -\frac{3}{2}\frac{1}{r}\frac{\partial T}{\partial r}B_{p0}\mathrm{sin}{\theta}\hat{e}_{\theta}\;.
\eeq

We can further estimate the partial derivatives of the surface temperature at 
the polar cap. For the radial derivative $\frac{\partial T}{\partial r}$ we use
the fact that all the heat flux arriving at the surface is radiated away as 
blackbody radiation.
\beq
\frac{\partial T}{\partial r}=-\frac{1}{\kappa}\sigma_{SB}T_s^4\;,
\label{eq:dTdr}
\eeq
where $\sigma_{SB}$ is the Stefan-Boltzmann constant and $\kappa$ is the heat 
conductivity. For $\kappa\approx 5\times 10^{13}$ erg s$^{-1}$ cm$^{-1}$ 
K$^{-1}$ and $T=5\times 10^6$ K, this results in 
$\frac{\partial T}{\partial r}\approx 7\times 10^8$ K cm$^{-1}$. The meridional
temperature gradient at the polar cap surface 
$\frac{\partial T}{R\partial \theta}$ corresponds to a drop of the temperature 
by a few $10^6$ K over the rim of the polar cap, i.e. over a distance of a few 
$10$ to about $100$ m. Therefore, $\frac{\partial T}{R\partial \theta}\gtrsim 
10^4$ K cm$^{-1}$. Hence, the radial term dominates the partial derivative of 
the surface temperature gradient. Also, in Eq.~\ref{eq:mdash-r} and 
\ref{eq:mdash-theta}, the second derivatives of the temperature can be safely 
neglected since they are much smaller than the first derivatives. Using these 
arguments we find 
\beq
m'=\vec{\nabla}(\vec{\nabla} T\cdot \vec{B})=\frac{3B_{p0}\sigma_{SB}T_s^4}{2\kappa R_{pc}}\mathrm{sin}{\theta}\vec{e}_{\theta}\;.
\label{eq:mdash}
\eeq
at the surface of the polar cap. Inserting the abbreviations we find the 
thermal drift oscillation frequency of the magnetic field perturbations as
\beq
\omega_{TD}=\frac{\eta_B k^2}{2}\sqrt{\frac{6\omega_B\tau c Q \xi \sigma_{SB} T_s^4}{B_{p0} \kappa R_{pc} \eta_B k^2}\mathrm{sin}{\theta}+1}\;.
\label{eq:omega_TD}
\eeq

\subsection{Rough estimates}
In order to get an idea of the magnitude of Hall and thermal magnetic field 
oscillations we consider the conditions at the polar cap surface with the local
dipolar poloidal background magnetic field. We consider two different field 
strengths $B_{p0}=10^{14}$ G and $B_{p0}=5\times 10^{13}$ G, covering roughly 
the range of inferred field strengths of polar caps \citep[see Table 1 
in][]{G17}. These field strengths imply that for Iron nuclei the zero pressure 
densities are $\rho_s=1.25\times 10^6$ and $5.1\times 10^5$ g cm$^{-3}$, 
respectively. We assume a surface temperature $T=T_s=5\times 10^6$ K. Using 
these values the electron number density $n_e$, the Fermi momentum $p_F$, and 
the effective electron mass $m_{\ast}$ are:
\beq
 n_e=\frac{\rho Z}{A m_u}\;, p_F=\hbar(3\pi^2n_e)^{1/3}\;,m_{\ast}=c^{-2}(m_e^2c^4+c^2p_F^2)^{1/2}\;
\eeq

To determine the coefficient $\xi$ we follow \cite{UY80}. As the condition 
$\Theta_D/4 < T<T_{melt}$ is clearly valid for surface temperatures $T\gtrsim 
3\times 10^6$ K ($\Theta_D$ and $T_{melt}$ are the Debye and melting 
temperature of the crystalline crust, respectively), this can be used to 
obtain: 
\beq
\xi=\frac{2-3\beta^2}{2+\beta^2},\;\beta=\frac{v_F}{c},\;v_F=\frac{p_F}{m_{\ast}}\;.
\label{eq:xi}
\eeq

The transport coefficients depend on the strong background field $B_{p0}$. Both
the perpendicular and `Hall' to $\vec{B}_{p0}$ components of the electric 
conductivity $\hat{\sigma}$, the heat conductivity $\hat{\kappa}$, and the `Hall' component
of the thermopower $\hat{Q}$ are drastically reduced by a factor $(\omega_B\tau)^2$. 
In case of the Hall drift induced oscillations this effect is automatically 
taken into account because the Hall-term in Eq.~\ref{eq:HallIndEq} arises from 
the tensor properties of $\hat{\sigma}$. The same is true for the thermal drift
term in Eq.~\ref{eq:E_T3}. The heat flux, however, dominates parallel to 
$\vec{B}_{p0}$ in the polar cap region, so that $\kappa$ appearing in 
Eq.~\ref{eq:dTdr} is weakly dependent on the local magnetic field. Therefore, 
we have used the scalar values of the transport coefficients that are parallel 
to $\vec{B}_{p0}$. The conductivities decrease with increasing field strength 
for constant density. In the density and temperature ranges under 
consideration, magnetic fields stronger than $\sim 10^{12}$G quantize, 
resulting in Shubnikov - de Haas oscillations of the conductivities 
\citep{SdH30}. The (transport-) coefficients were calculated using publicly 
available codes provided by Potekhin et al. (see http://www.ioffe.ru/astro/EIP/index.html and http://www.ioffe.ru/astro/conduct). For the given density, 
temperature, and using Iron nuclei as constituent of the polar cap surface the 
estimated transport coefficients are shown in Tab.~\ref{tab:coefficients} (the
impurity content of the crystalline layer has minimal effect due to the 
relatively low density and high temperature).

\begin{table*}
    \caption{Transport coefficients.
     }
    \label{tab:coefficients}
    \begin{center}
   \begin{tabular}{lccccccccc}
        \hline
        \hline
           $B_{p0}$   &   $\rho_s$   &   $n_e$   &   $m_{\ast}$   &   $\xi$   &  $\sigma$  &  $\kappa$  &  $Q$  &  $\omega_B\tau$   \\
         (G)  & (g cm$^{-3}$) &  (cm$^{-3}$) & (g) &   & (s$^{-1}$) & (g cm K$^{-1}$ s$^{-3}$)  & (G cm K$^{-1}$) &   \\ 
       \hline
&&&&&&&&& \\
$10^{14}$ & $1.25\times 10^{6}$ & $3.5\times 10^{29}$ & $1.2\times 10^{-27}$ & $0.32$ & $5.5\times 10^{19}$ & $7.5\times 10^{13}$ & $6.1\times 10^{-8}$ & $1.1\times 10^3$ \\
$5\times 10^{13}$ &$5.1\times 10^{5}$ & $1.4\times 10^{29}$ & $1.1\times 10^{-27}$ & $0.51$ & $2.1\times 10^{19}$ & $2.8\times 10^{13}$ & $9.3\times 10^{-8}$ & $5.2\times 10^2$ \\
&&&&&&&&&\\
\hline
\end{tabular}
\end{center}
\end{table*}

The frequencies of the Hall and thermal drift oscillations are calculated using
the transport coefficients for two values of $B_{p0}, T_s=5\times 10^6$ K, 
$\theta=45^{\circ}$, and $R_{pc}=5\times 10^3$ cm. Both $\omega_{Hall}$ and 
$\omega_{TD}$ depend strongly on the wave vector of the magnetic perturbations,
which may vary over a relatively wide range. As a result we have only checked 
some extreme cases. The biggest scale is perhaps the radius of the polar cap 
$R_{pc}$ which vary between $\sim 10$ m and few $100$ m \citep[see Table 1 
in][]{G17}. In order to get a qualitative idea about the oscillation 
frequencies we consider two cases, $k=k_r=k_{\theta}= 6.2\times 10^{-4}$ and 
$6.2\times 10^{-3}$ cm$^{-1}$, which correspond to $R_{pc}\approx 100$ and $10$
m, respectively. Using these parameters we have estimated both the Hall and 
thermal drift oscillation frequencies of the magnetic field perturbations given
in Eqs.~\ref{eq:DispHall} and \ref{eq:omega_TD}. The oscillation periods and 
the characteristic Ohmic decay times ($T_{Ohm}=1/\eta_B k^2$) are reported in 
Tab.~\ref{tab:results}.

\begin{table*}
    \caption{Ohmic decay time, Hall and thermal drift oscillation periods for  typical wavelengths and surface magnetic field strengths.
     }
    \label{tab:results}
    \begin{center}
   \begin{tabular}{lcccccc}
        \hline
        \hline
           $B_{p0}$   &   $k$   &   $T_{Ohm}$   &   $T_{Hall}$   &   $T_{TD}$   &  $\eta_B$ \\
         (G)  & (cm$^{-1}$) & (s) & (s) & (s) & (cm$^2$ s$^{-1}$) \\
       \hline
&&&&&& \\
$10^{14}$ & $6.2\times 10^{-3}$ & $2\times 10^4$ & $120$ & $3.4\times 10^4$ & $1.3$ \\
$10^{14}$ & $6.2\times 10^{-4}$ & $2\times 10^6$ & $1.2\times 10^4$ & $3.5\times 10^5$ & $1.3$ \\
$5\times 10^{13}$ & $6.2\times 10^{-3}$ & $ 7.7\times 10^3$ & $90$ & $8.6\times10^3$ & $3.4$ \\
$5\times 10^{13}$ & $6.2\times 10^{-4}$ & $7.7\times 10^5$ & $9\times 10^3$ & $8.7\times10^4$ & $3.4$ \\

&&&&&&\\
\hline
\end{tabular}
\end{center}
\end{table*}

\section{A model for the short timescale phenomena in pulsars}
\label{sec3}
In Tab.~\ref{tab:results} the oscillation periodicities corresponding to the 
Hall and thermal drifts show similarity with the the timescales associated with
ST phenomena (see Fig.~\ref{fig1}). This opens up the possibility that the 
variations on the polar cap surface magnetic field due to the Hall and thermal 
drifts are responsible for some of the ST phenomena seen in pulsars. However,
these variations alone are not sufficient to explain the observed behaviour 
during the state changes. For example the transitions between the different 
states happen very rapidly, usually within a period. In some cases the X-ray 
emission shows synchronous variations during the transition to different modes,
while in others the subpulse drift show different behaviour in each mode. In 
order to explain some of these effects associated with the ST phenomena we 
propose a model based on the variations of the sparking process in the 
partially screened gap (PSG) above the non-dipolar polar cap.

The PSG model considers a steady flow of iron ions from the stellar surface 
which screens the accelerating electric field of the gap by a screening factor
$\eta = \left(1-n_{i}/n_{GJ}\right)$, here $n_{i}$ is the ion number 
density and $n_{GJ}$ is the Goldreich-Julian density. Positive charges from the
stellar surface cannot be supplied at a rate that would compensate the inertial
outflow through the light cylinder. As a result, significant potential drop 
develops above the polar cap, and consequently the gap breaks down by creating 
electron-positron pairs which setup sparking discharges in the gap. The charges
are accelerated in opposite dictions to relativistic speeds by the very high 
electric fields and and back-flowing electrons heats the surface to temperature
$T_{s} > 10^{6}$ K. Thermal ejection of iron ions causes a partial screening of
the acceleration potential drop and the backflow heating decreases as well. 
Thus heating leads to cooling, which resembles a classical thermostat. Surface 
temperature is thermostatically regulated to retain its value close to critical
temperature $T_{0}$ above which thermal ion flow reaches the Goldreich-Julian 
densities - $T_{s}\lesssim T_{0}$. According to calculations of cohesive energy
by \citet{ML07}, this can occur if the surface magnetic field $B$ is close to 
$10^{14}$ G, which is indicative of highly non-dipolar field. This also implies
that the actual polar cap area, $A_{pc}$, is smaller than the canonical dipolar
polar cap, $A_{0}$, by the factor $b=B/B_{0}=A_{0}/A_{pc}$. Here $B_{0}=2\times10^{12}\left(\dot{P}_{-15}P\right)^{0.5}~$G, $A_{0}=$ $6.6\times10^{8}P^{-1}$ 
cm$^{2}$, $P$ is the pulsar period and $\dot{P}_{-15}=10^{15}\dot{P}$ is the 
period derivative. 

It is further required that in order to support the PSG, the polar cap at any 
given time should be packed with sufficient number of sparks that provide the 
necessary heating of the stellar surface below the spark, and generate a pair 
plasma cloud moving outward along the open field lines. The subpulse associated
observed radio emission should be generated in these plasma clouds consisting 
of relativistic electrons and positrons as well as iron ions. The observed 
subpulse drifting phenomenon is explained by the charges lagging behind the 
corotation motion (\citealt{2020MNRAS.496..465B}). Thus the shape of pulsar 
profiles and subpulse drifting characteristics should be defined by the spark 
behavior in the PSG. It is important to underline that the basic 
characteristics of spark, i.e. their size and drifting rate can be estimated in
the PSG model without any further assumptions. We can estimate the number of 
sparks populating the polar cap as
\[
N_{sp}=4\times10^{2}\left(\dfrac{B_{14}}{T_{6}^{4}}\right)\left(\dfrac{\dot{P}_{-15}}{P^{5}}\right)^{0.5}\left(\dfrac{\eta}{0.1}\right)^{2}.
\] 
Here $B_{14}=10^{-14}B$ and $T_{6}=10^{-6}T_{s}$. According to \cite{ML07} if 
$B_{14}=1$, then $T_{6}=2$, and if $B_{14}=0.5$, then $T_{6}=1$. As an example 
we can assume $P\approx$ $\dot{P}_{-15}\approx1$ and $\eta \approx0.1$, and 
then we obtain $N_{sp}\approx25$ in the first case and $N_{sp}\approx200$ in 
the second case. Based on the PSG model we can also estimate the subpulse drift
periodicity, $P_{3}$, in pulsars showing systematic drift motion. $P_{3}$ is 
the interval between the signal repeating at the same location and it is used 
to estimate the drift velocity of sparks in the polar cap region. In the PSG 
model we can estimate $P_{3}$ as \citep[see][]{BMM16}.
\[
\left(\dfrac{P_{3}}{P}\right) = 2.25~\left(\dfrac{\eta}{0.1}\right)^{-1}.
\]
From the PSG model it also follows that pulsars' X-ray luminosity ($L_{X}$) is 
proportional to the screening factor, $L_{X}\sim\eta$ \citep{BMM16}. Thus, we 
demonstrated that most phenomena which show variations during mode changing are
affected by the parameter $\eta$ (see Tab.~\ref{tabmode}), as changing of 
$N_{sp}$ should also result in the change of average profile and other emission
characteristics.

We now propose a qualitative model for mode changing based on oscillations of 
magnetic field perturbations caused by either the Hall drift (short timescales) 
or the thermal drift (longer timescales). The curvature ($\rho_{c}$) of the 
magnetic field lines is highly sensitive to alteration of the surface magnetic 
field \citep[Fig. 8 in][]{GMM02}. On the other hand changing of $\rho_{c}$ 
strongly affects the pair creation process as well as alters the gap closing 
timescales and average Lorentz factors of the secondary plasma particles. As a 
result any change in $\rho_{c}$ directly affects $\eta.$ Therefore, we can 
speculate, that during the magnetic field perturbation oscillation, when a 
cumulative change of $\eta$ is high enough to cause a change of $N_{sp}$ and 
consequently to cause rearrangement of a spark distribution, the mode changing 
is induced. The alternative mode exists as long as the perturbation is in the 
phase near to the maximum deviation. The sudden transitions between the modes 
as well as their unique identities also find ready explanation in this 
mechanism. This model requires that mode changing pulsars should have a 
moderate number of sparks, otherwise it would be difficult to observe the 
alteration of average profiles.

In case of pulsars that have a larger number of sparks (say more than 10) on 
their polar caps, the change of number of sparks by one, is unlikely to modify 
the profile shape. But oscillating magnetic field perturbations can alter the 
curvature of the magnetic field lines and change the screening factor $\eta$ in 
the pair creation region, during the change in the number of sparks. As a 
result, the parameters of the secondary plasma clouds, that generate the radio 
emission, may be modified. As shown in \citet{GLM04} the total radio luminosity
($L$) of pulsars are very sensitive to the Lorentz factors of the secondary 
plasma ($\gamma$) and the bunch particles ($\Gamma$). This dependence can be 
expressed as \citep[see eq.47 in][]{GLM04}:
\be
L\propto \left( \dfrac{\Gamma^{12}}{\gamma^{3.5}}\right) 
\ee
Hence, this may lead to not only amplitude modulations but also periodic 
nulling where the emission intensity goes below detection limit due to change 
in the plasma parameters.

Thus, two types of alteration of the pulsar radio emission can be distinguished
in this proposed mechanism. The first type corresponds to pulsars with a 
moderate number of sparks and manifests itself as mode changing. The 
distribution of sparks along the observer's line of sight changes enough to 
cause the profile shape to change. These changes may also be accompanied by 
change in subpulse drifting behaviour and X-ray emission properties. The second
type is expected in pulsars with a higher number of sparks. In this case, 
changing the number of sparks by one changes the spark size slightly, which is 
unlikely to modify the profile shape significantly, but nevertheless changes 
the secondary plasma parameters such as the values of the characteristic 
Lorentz factors. Therefore, periodic amplitude modulation as well as periodic 
nulling can be expected in this case.

The oscillation frequencies of local magnetic field perturbations at the polar
cap surface driven by the Hall drift and thermoelectric effects are calculated 
based on linear dispersion analysis. It assumes that these perturbations are 
sufficiently small in comparison to the background magnetic field. In case this
assumption is not valid, i.e. the perturbations exceed 
$(\delta B/B_{p0})^2 \sim 1$, nonlinear effects become important. Then, a 
non-linear dispersion analysis has to be performed. It will reveal interactions
of different modes of the local field structure and will describe deviations 
from the here presented purely oscillatory behaviour of the local field 
structures as given by $T_{Hall}$ and $T_{TD}$.

\section{Conclusion}
\label{sec4}
The combined Hall and thermal drift causes oscillations of the local magnetic field
structure at the surface of the polar cap of radio pulsars. As seen in  Table ~\ref{tab:results}, 
the period of Hall oscillations is for typical wave vectors  much smaller than the corresponding 
Ohmic decay times. The same is true for thermal drift oscillations with wave vectors
$\lesssim 10^{-1}$ cm$^{-1}$, correspnding to wave lengths $\lambda \gtrsim 60$m.
 Magnetic perturbations of such wavelengths fit very well into the diameter of the polar cap.
 Hence, both the Hall and thermal drift driven oscillations of the local magnetic field structure will
 not be affected by Ohmic diffusion.
 Hall and thermal drift oscillations have very different timescales. 
The shortest oscillation periods are of the order of a few minutes and caused 
by Hall drift waves with wavelengths of $\sim 10 - 100$ m. The thermal drift,
combined to the Hall drift, 
drives magnetic field oscillations of the same wavelengths over timescales of 
few hours. The radio emission shows several short timescale phenomena like mode
changing, periodic amplitude modulations and periodic nulling, which also show 
changes in the emission states over timescales of few minutes to several hours.
We propose that these state changes can be realised using the PSG model, where 
the perturbations in the magnetic field due to Hall and thermal drift 
oscillations changes the sparking configuration. In this model the primary 
difference between mode changing and the other periodic modulations is the 
number of sparks in the PSG. Mode changing is seen when there are fewer sparks,
while the periodic amplitude modulation and periodic nulling requires a larger 
number ($>$10). However, our analytical analysis is not sufficient to explain 
several observational features in greater detail, e.g., the observed 
quasi-periodic nature of the transitions. A more detailed numerical approach is
required to address several issues, like non-linear dispersion analysis, 
relaxing the approximations of axial symmetry and independence of the transport coefficients 
of coordinates, etc., which will be taken up in future works.

\section*{Acknowledgement}
We thank the referee for comments that improved the paper.
This work was supported by the grant 2020/37/B/ST9/02215 of the National 
Science Centre, Poland. DM acknowledges the support of the Department of Atomic
Energy, Government of India, under project No. 12-R\&D-TFR-5.02-0700. DM 
acknowledges funding from the grant ``Indo-French Centre for the Promotion of 
Advanced Research—CEFIPRA" grant IFC/F5904-B/2018.

\section{Appendix}
\subsection {The Mode changing phenomenon in the pulsar population}
In this section of the appendix we have collated the known mode changing 
behaviour in pulsars from the literature, where detailed modal timescales are 
available. Table \ref{tabmode} lists the different emission modes in each 
pulsar, along with the typical modal durations, the percentage of time the 
pulsar spends in each mode and the reference study. The moding behaviour is 
divided into four different groups. The first group includes 14 pulsars with 
traditional mode changing in the form of distinct profile shapes in different 
modes. The second group of 9 pulsars is associated with subpulse drifting, 
where the drifting behaviour changes in the different modes. In the third group
there are 5 pulsars where the pulsar transitions from a normal emission state 
to a bursting state resembling RRAT emission. Finally, there are three pulsars 
where the emission switches to a flaring state preceding the pulse window, in a
quasi-periodic manner.

\begin{table*}
\caption{The mode changing phenomenon seen in Pulsars.}
\label{tabmode}
\centering
\begin{tabular}{ccccccc}
\hline
  & Pulsar & Period & Mode & Typical Length & Abundance & Reference \\
  &        &  (sec) &      &             &           &           \\
\hline
\multicolumn{7}{c}{\underline{\large Mode changing characterised by Profile change}} \\
   &  &  &  &  &  &  \\
 1 &  B0329+54  & 0.715 & Normal & $\sim$150 mins & $\sim$85\% & 1, 2, 3 \\
   &  &  & Abnormal - A, B, C & $\sim$30 mins & $\sim$15\% & \\
   &  &  &  &  &  &  \\
 2 &  B0355+54  & 0.156 & Normal & --- & --- & 4 \\
   &  &  & Abnormal & 3600$P$ & $\leq$5\% &      \\
   &  &  &  &  &  &  \\
 3 &  B0823+26  & 0.531 & B & $\sim$ few hrs & --- & 5, 6 \\
   &  &  & Q & $\sim$ few hrs & --- &  \\
   &  &  &  &  &  &  \\
 4 &  B0844-35  & 1.116 & A & 10-15 mins & --- & 7 \\
   &  &  & B & $\sim$10 secs & --- &  \\
   &  &  &  &  &  &  \\
 5 & J1107-5907 & 0.253 & Strong & 200-6000$P$ & $\sim$8\% & 8, 9 \\
   &  &  & Weak/Apparent Null & --- & --- &      \\
   &  &  &  &  &  &  \\
 6 &  B1237+25  & 1.382 & Normal-quiet/flare & 60-80 $P$ cycle & --- & 10, 11 \\
   &  &  & Abnormal & few $P$ to several 100 $P$ & --- &  \\
   &  &  &  &  &  &  \\
 7 &  B1358-63  & 0.843 & 2 Modes & $\sim$100 $P$ & --- & 12 \\
   &  &  &  &  &  &  \\
 8 & J1658-4306 & 1.166 & A & 10-20 mins & --- & 7 \\
   &  &  & B & 3-4 mins & $\sim$20\% &  \\
   &  &  &  &  &  &  \\
 9 &  B1658-37  & 2.455 & A & --- & --- & 7 \\
   &  &  & B & 1-2 mins & $\sim$10\% &  \\
   &  &  &  &  &  &  \\
10 & J1703-4851 & 1.396 & A & few to several 10 mins & --- & 7 \\
   &  &  & B & few secs to 1-2 mins & $\sim$15\% &  \\
   &  &  &  &  &  &  \\
11 &  B1822-09  & 0.769 & B & $\sim$200 $P$ & --- & 13, 14, 15, 16, 17, 18 \\
   &  &  & Q & $\sim$350 $P$ & $\sim$64\% &  \\
   &  &  &  &  &  &  \\
12 & J1843-0211 & 2.028 & A & few mins to few 10 mins & --- & 7 \\
   &  &  & B & few mins & few \% &  \\
   &  &  &  &  &  &  \\
13 &  B1926+18  & 1.220 & Normal & 500-1000 $P$ & 85-90\% & 19 \\
   &  &  & Abnormal & 200-300 $P$ & --- &  \\
   &  &  &  &  &  &  \\
14 &  B2020+28  & 0.343 & Normal & --- & 89\% & 20 \\
   &  &  & Abnormal & $<$ 250 $P$ & 11\% &  \\
   &  &  &  &  &  &  \\
%   &  &  &  &  &  &  \\
\multicolumn{7}{c}{\underline{\large Mode changing characterised by change in Subpulse Drifting}} \\
   &  &  &  &  &  &  \\
 1 &  B0031-07  & 0.943 & A & $\sim$55 $P$ & $\sim$10-20\% & 21, 22, 23, 24 \\
   &  &  & B & $\sim$30 $P$ & $\sim$35\% &  \\
   &  &  & C & $\sim$11 $P$ & $\sim$1\% &  \\
   &  &  & Null & $\sim$32 $P$ & $\sim$45\% &  \\
   &  &  &  &  &  &  \\
 2 &  B0809+74 & 1.292 & Normal & $\sim$500-1000 $P$ & --- & 25 \\
   &  &  & Slow Drift & $\sim$100 $P$  & --- &  \\
   &  &  &  &  &  &  \\
 3 &  B0943+10  & 1.098 & B & $\sim$7.5 hrs & 77\% & 16, 26, 27, 28, 29, 30, 31 \\
   &  &  & Q & $\sim$2.2 hrs & 23\% &  \\
   &  &  &  &  &  &  \\
 4 &  B1819-22  & 1.874 & A & 82 $P$ & 45\% & 32 \\
   &  &  & B & 68 $P$ & 38\% &  \\
   &  &  & C & $\sim$200 $P$ & $<$4\% &  \\
   &  &  &  &  &  &  \\
 5 &  B1918+19  & 0.821 & A & 35 $P$ & 17\% & 33, 34 \\
   &  &  & B & 53 $P$ & 48\% &  \\
\hline
\end{tabular}
\end{table*}

\begin{table*}
\contcaption{The mode changing phenomenon seen in Pulsars.}
\label{tabmode:cont}
\centering
\begin{tabular}{ccccccc}
\hline
  & Pulsar & Period & Mode & Typical Length & Abundance & Reference \\
  &        &  (sec) &      &             &           &           \\
\hline
   &  &  & C & 135 $P$ & 14\% &  \\
   &  &  & N & 23.5 $P$ & 21\% &  \\
   &  &  &  &  &  &  \\
 6 &  B1944+17  & 0.441 & A & 30-40 $P$ & $\sim$7\% & 35, 36 \\
   &  &  & B & 10-20 $P$ & $\sim$2\% &  \\
   &  &  & C & 8$P$ & $\sim$13\% &  \\
   &  &  & D & 11-12$P$ & $\sim$12\% &  \\
   &  &  & Null & 20-100 $P$ & $\sim$66\% &  \\
   &  &  &  &  &  &  \\
 7 &  B2003--08 & 0.580 & A & 65 $P$ & 15\% & 37 \\
   &  &  & B & 103 $P$ & 13\% &  \\
   &  &  & C & 16 $P$ & 7\% &  \\
   &  &  & D & 15 $P$ & 22\% &  \\
   &  &  & Null & $\sim$100 $P$ & 29\% &  \\
   &  &  &  &  &  &  \\
 8 &  B2303+30  & 1.576 & B & 37 $P$ & $\sim$54\% & 38 \\
   &  &  & Q & 31 $P$ & $\sim$46\% &  \\
   &  &  &  &  &  &  \\
 9 &  B2319+60  & 2.256 & A & 30-70 $P$ & 30-45\% & 39, 40 \\
   &  &  & B & 12-15 $P$ & 5-15\% &  \\
   &  &  & ABN & 10-15 $P$ & 5-10\% &  \\
   &  &  & C & 20-30 $P$ & 10-20\% &  \\
   &  &  & Null & $\sim$10 $P$ & $\sim$35\% &  \\
%   &  &  &  &  &  &  \\
   &  &  &  &  &  &  \\
\multicolumn{7}{c}{\underline{\large Mode changing characterised by Bursting Emission}} \\
   &  &  &  &  &  &  \\
 1 &  B0611+22  & 0.335 & Normal & $\sim$1200 $P$ & --- & 41, 42 \\
   &  &  & Burst & 300-600 $P$ & --- &  \\
   &  &  &  &  &  &  \\
 2 &  B0826-34  & 1.849 & Strong/Normal & $\sim$few hrs & --- & 43, 44, 45 \\
   &  &  & Weak/RRAT type & $\sim$few hrs & $\sim$70\% &  \\
   &  &  &  &  &  &  \\
 3 &  J0941-39  & 0.587 & On & --- & --- & 46 \\
   &  &  & Burst/RRAT & hrs to even weeks & --- &  \\
   &  &  &  &  &  &  \\
 4 & J1752+2359 & 0.409 & Bright & $\sim$88 $P$ & --- & 47, 48 \\
   &  &  & Bursts within Null & $\sim$570 $P$ & $<$89\% &  \\
   &  &  &  &  &  &  \\
 5 & J1938+2213 & 0.166 & Normal & --- & --- & 49 \\
   &  &  & Bursts & 20-25 $P$ & $\sim$1\% &  \\
%   &  &  &  &  &  &  \\
   &  &  &  &  &  &  \\
\multicolumn{7}{c}{\underline{\large Mode changing characterised by Preceding Flare}} \\
   &  &  &  &  &  &  \\
 1 &  B0919+06  & 0.431 & Normal & 1000-3000 $P$ & --- & 50, 51, 52, 53 \\
   &  &  & Preceding Flare & few 10 $P$ & $\sim$2\% &  \\
   &  &  &  &  &  &  \\
 2 &  B1322-66  & 0.543 & A & 200-1000 $P$ & --- & 7, 54 \\
   &  &  & B/Preceding component & $<$ 100 $P$ & --- &  \\
   &  &  &  &  &  &  \\
 3 &  B1859+07  & 0.644 & Normal & $\sim$150 $P$ & --- & 50, 53, 55 \\
   &  &  & Preceding Flare & few $P$ to few 10 $P$ & $\sim$20\% &  \\
\hline
\end{tabular}
\medskip
\\1-\cite{L71}; 2-\cite{BMS82}; 3-\cite{CWW11}; 4-\cite{MSF82}; 5-\cite{SYH15};
6-\cite{BM19}, 7-\cite{WMJ07}; 8-\cite{YWS14}; 9-\cite{WHK20}; 10-\cite{B70b}; 
11-\cite{SR05}; 12-\cite{VDH97}; 13-\cite{FMW81}; 14-\cite{FW82}; 
15-\cite{GJK94}; 16-\cite{BMR10}; 17-\cite{LMR12}; 18-\cite{HKH17}; 
19-\cite{FBW81}; 20-\cite{WWY16}; 21-\cite{HTT70}; 22-\cite{VJ97}; 
23-\cite{SMK05}; 24-\cite{MBT17}; 25-\citep{vLSRR03}; 26-\cite{SIR98}; 
27-\cite{RSD03}; 28-\cite{RS06}; 29-\cite{SR09}; 30-\cite{BMR11}; 
31-\cite{H13}; 32-\cite{BM18b}; 33-\cite{HW87}; 34-\cite{RWB13}; 
35-\cite{DCH86}; 36-\cite{KR10}; 37-\cite{BPM19}; 38-\cite{RWR05}; 
39-\cite{WF81}; 40-\cite{RBMM21}; 41-\cite{SLR14}; 42-\cite{RSL16}; 
43-\cite{ELG05}; 44-\cite{S11}; 45-\cite{EAM12}; 46-\cite{BB10}; 
47-\cite{LWF04}; 48-\cite{GJW14}; 49-\cite{LCM13}; 50-\cite{RRW06}; 
51-\cite{PSW15}; 52-\cite{HHP16}; 53-\cite{WOR16}; 54-\cite{WYY20}; 
55-\cite{PSW16};
\end{table*}

%%%%%%%%%%%%%%%%%%%%%%%%%%%%%%%%%%%%%%%%%%%%%%%%%%%%%%%%%%%%
\subsection {Units}
%%%%%%%%%%%%%%%%%%%%%%%%%%%%%%%%%%%%%%%%%%%%%%%%%%%%%%%%%%%%

%$[\eta_B]=cm^2s^{-1}$, $[\alpha_H]=G^{-1}$, $[\alpha]=cm^2s^{-1}G^{-1}$,\\
$[\eta_B]=cm^2s^{-1}$, $[\alpha_H]=cm^2s^{-1}G^{-1}$,\\
$[\beta_T]=cm^2s^{-1}G^{-1}K^{-1}$, $[m]=Gcm^{-1}K$, $[m']=Gcm^{-2}K$, $[n]=cms^{-1}$.\\
$[Q]=g^{1/2}cm^{1/2}s^{-1}$K$^{-1}=GcmK^{-1}$

%%%%%%%%%%%%%%%%%%%%%%%%%%%%%%%%%%%%%%%%%%%%%%%%%%%%%%%%%%%%
\subsection{Fourier transformations in detail}
\label{sec:FT}
%%%%%%%%%%%%%%%%%%%%%%%%%%%%%%%%%%%%%%%%%%%%%%%%%%%%%%%%%%%%

\noindent Fourier transformations return for the temporal and spatial partial derivatives the following expressions:

\beq
\frac{\partial \delta\vec{ B}_{p,t}}{\partial t} =i\omega\delta\vec{B}_{p,t}\mathrm{exp}\left[i(\omega t - \vec{k}\cdot\vec{r})\right]\;,
\nonumber \\
\vec{\nabla}\times\delta\vec{B}_{p,t}=-i\vec{k}\times  \delta\vec{B}_{p,t}\mathrm{exp}\left[i(\omega t - \vec{k}\cdot\vec{r})\right]\;.
\label{eq:FT}
\eeq

\noindent With 

\beq 
\vec{\nabla}\times\left(\vec{\nabla}\times \delta\vec{ B}_{p,t}\mathrm{exp}\left[i(\omega t - \vec{k}\cdot\vec{r})\right]\right)
\nonumber\\
=\vec{\nabla}\times\left[-i\vec{k}\times\delta\vec{B}_{p,t}\mathrm{exp}\left[i(\omega t - \vec{k}\cdot\vec{r})\right]\right]
\nonumber\\
= -(-i\vec{k}(-i\vec{k})\delta\vec{ B}_{p,t} \mathrm{exp}\left[i(\omega t - \vec{k}\cdot\vec{r})\right]\;,
\label{eq:curlcurl}
\eeq

\noindent we find for the pure Ohmic decay
\be
i\omega  \delta\vec{ B}_{p,t}=\frac{c^2k^2}{4\pi\sigma_0}\delta\vec{ B}_{p,t}\;,
\label{eq:Ohm}
\ee

\noindent which gives the well known dispersion relation for the Ohmic decay of a magnetic field with the Ohmic decay time $\tau_{Ohm}=\frac{4\pi\sigma_0}{c^2k^2}$. Identifying the wavelength $\lambda$ of theabsolute value of the wave vector $k=2\pi/\lambda$ with the radius of a conducting sphere, $\lambda=R$, we find the Ohmic decay time for the dipolar component of a magnetic field as $\tau_{Ohm,dip}=\sigma_0 R^2/\pi c^2$.

%%%%%%%%%%%%%%%%%%%%%%%%%%%%%%%%%%%%%%%%%%%%%%%%%%%%%%%%%
\subsection { Magnetization parameter $\omega_B\tau$ in detail}
%%%%%%%%%%%%%%%%%%%%%%%%%%%%%%%%%%%%%%%%%%%%%%%%%%%%%%%%%

\beq
\omega_B\tau=\frac{eB}{m_{\ast}c}\tau\;\;, \tau=\frac{m_{\ast}\sigma}{e^2n_e}\;\;, n_e=\frac{\rho Z}{A m_u}\;,
\eeq

\noindent where e.g at $\rho=1.25\times 10^6$ g cm$^{-3}$ and $T=5\times 10^6$ K the electric conductivity $\sigma\approx  5.5\times 10^{19}$ s$^{-1}$ and the electron number density $n_e\approx 3.5\times 10^{29}$ cm$^{-3}$. Thus we find for  $B_{p0}=10^{14}$ G a magnetization parameter of $\omega_B\tau=\frac{B\sigma}{cen_e}\approx 1.1\times 10^3$.

%%%%%%%%%%%%%%%%%%%%%%%%%%%%%%%%%%%%%%%%%%%%%%%%%%%%%%%%
\subsection{Derivation of the thermoelectric field (Eq.~\ref{eq:E_T})}
%%%%%%%%%%%%%%%%%%%%%%%%%%%%%%%%%%%%%%%%%%%%%%%%%%%%%%%%
\label{sec:8.4}

\noindent We follow \cite{UY80} to derive the equations which describe the thermal drift velocity of the magnetic field in presence of a temperature gradient. The electric field created by temperature gradients in the presence of magnetic fields is

\beq
\vec{E}_T=-Q_{\parallel}\vec{\nabla}_{\parallel}T-Q_{\perp}\vec{\nabla}_{\perp}T-Q_{\wedge}\vec{\nabla}_{\wedge}T\;
\label{eq:E_general}
\eeq

\noindent where $Q_{\parallel},\;Q_{\perp},\;Q_{\wedge}$ are tensor components of the thermopower parallel, perpendicular, and "Hall"' to the unit vector of the magnetic field $\vec{b}=\vec{B}/\mid{\vec{B}}\mid$. These tensor components are given by \citep{UY80}:

\beq
Q_{\parallel}&=&Q(3+\xi ),
\nonumber\\
Q_{\perp}&=&Q\left[3+\xi(1+\omega_B^2\tau^2)^{-1}\right]\;,
\nonumber\\
Q_{\wedge}&=&Q\xi\omega_B\tau(1+\omega_B^2\tau^2)^{-1}\;,
\label{eq:thermopower}
\eeq

\noindent With the expressions for the components of an arbitrary vector $\vec{A}$  parallel, perpendicular, and "Hall"' to the magnetic field

\be
\vec{A}_{\parallel}=\vec{b}(\vec{A}\cdot\vec{b}), \vec{A}_{\perp}=\vec{b}\times(\vec{A}\times\vec{b}), \vec{A}_{\wedge}=\vec{A}\times \vec{b},  \vec{b}=\vec{B}/\mid{\vec{B}}\mid \;,
\ee

\noindent the thermoelectric field given in Eq. ~\ref{eq:E_T} can be written as

\beq
\vec{E}_T&=&-Q(3+\xi)\vec{\nabla}_{\parallel}T
\nonumber\\
&-&Q[\left[3+\xi(1+\omega_B^2\tau^2)^{-1}\right]\left[\vec{b}\times(\vec{\nabla}T\times\vec{b}\right]
\nonumber\\
&-&Q\xi\omega_B\tau(1+\omega_B^2\tau^2)^{-1}(\vec{b}\times\vec{\nabla}T)\;,
\label{eq:E_T1}
\eeq

\noindent where the second and the third line of Eq. ~\ref{eq:E_T1} represent the perpendicular and  "Hall" to the magnetic field components of $\vec{E}_T$. In the next step we pull out the factor $(1+\omega_B^2\tau^2)^{-1}$ by

\beq
3+\xi(1+\omega_B^2\tau^2)^{-1}&=&3+\xi-\xi\left[1-(1+\omega_B^2\tau^2)^{-1}\right]
\nonumber\\
&=&3+\xi-\xi\frac{\omega_B^2\tau^2}{(1+\omega_B^2\tau^2)}\;.
\eeq

\noindent It is $\vec{\nabla}_{\parallel}T+\vec{\nabla}_{\perp}T=\vec{\nabla}T$, i.e. the terms belonging to $\vec{\nabla}T$ can be combined with $3+\xi$:

\beq
\vec{E}_T&=&-Q(3+\xi)\vec{\nabla}T
\nonumber\\
&+&\frac{Q\xi}{1+\omega_B^2\tau^2}\bigl[(1+\omega_B^2\tau^2-1)\left(\vec{b}\times(\vec{\nabla}T\times\vec{b})\right)
\nonumber\\
&-&\omega_B\tau(\vec{b}\times\vec{\nabla}T)\bigr]
\nonumber\\
&=&-Q(3+\xi)\vec{\nabla}T
\nonumber\\
&+&\frac{Q\xi\omega_B\tau}{1+\omega_B^2\tau^2}\left[\vec{\nabla}T+\omega_b\tau(\vec{b}\times\vec{\nabla}T)\right]\times \vec{b}\;.
\label{eq:E_T2}
\eeq

\noindent The term on the last line of Eq. ~\ref{eq:E_T2} can be interpreted, similar to the Hall drift , as the thermal drift of the magnetic field with the thermal drift velocity, $\vec{v}_{TD}/c\times \vec{B}$ and

\be
\vec{v}_{TD}=\frac{Q\xi e\tau}{m_{\ast}(1+\omega_B^2\tau^2)}\left[\vec{\nabla}T+\omega_B\tau(\vec{b}\times\vec{\nabla}T)\right]\;.
\label{eq:v_TD}
\ee

\noindent In the limit $\omega_B\tau\gg1$, which is surely realized in the polar cap region of radio pulsars, $\vec{E}_T$ simplifies to

\be
\vec{E}_T=-Q(3+\xi)\vec{\nabla}T + Q\xi(\vec{b}\times\vec{\nabla}T)\times \vec{b}.
\label{eq:E_T_simpl}
\ee

%%%%%%%%%%%%%%%%%%%%%%%%%%%%%%%%%%%%%%%%%%%%%%%%%%%%%%%%%
\subsection{Derivation $\omega_{+}$}
\label{sec:compl_number}
%%%%%%%%%%%%%%%%%%%%%%%%%%%%%%%%%%%%%%%%%%%%%%%%%%%%%%%%

\noindent  Evaluating the square root in Eq. ~\ref{eq:DispTD} we use the representation of the square root of a complex number

\be
(a+ib)^{1/2}=(a^2+b^2)^{1/4}\times
\nonumber\\
\ee
\be
\times \left[\mathrm{cos}\left(\frac{1}{2}\mathrm{arctg}(b/a)\right)+i\mathrm{sin}\left(\frac{1}{2}\mathrm{arctg}(b/a)\right)\right]\;.
\label{eq:comproot}
\ee

\noindent In case $b/a<1$ is $\mathrm{arctg}(b/a)\lesssim1$ and, hence, $\mathrm{cos}(\frac{1}{2}\mathrm{arctg}(b/a)  \approx \mathrm{cos}(1/2) \approx 0.9$, where  $a^2=(\eta_B^2 k^4/4+n\beta_T(m'\cdot\vec{k}))^2,\;\;b^2=n^2\beta_T^2m^2k^4$.  If $b/a <1$, i.e. $b^2/a^2\ll 1$:

\beq
\frac{b^2 }{a^2}=\frac{n^2\beta_T^2m^2k^4}{\left(n\beta_T(m'\cdot\vec{k})+\frac{\eta_B^2 k^4}{4}\right)^2}\ll1
\eeq

\noindent follows for the real part of $\omega_{\pm}$

\beq
\omega_+\approx\frac{\sqrt{a}}{2}=\frac{\eta_B k^2}{2}\sqrt{\frac{4n\beta_T(m'\cdot\vec{k})}{\eta_B^2k^4}+1}\;.
\label{eq:omega_plus}
\eeq

\section*{Data availability}
No new data were generated or analysed in support of this research.
%%%%%%%%%%%%%%%%%%%%%%%%%%%%%%%%%%%%%%%%%%%%%%%%%%%%%%%%%%%%
\bibliography{pulsars}
%%%%%%%%%%%%%%%%%%%%%%%%%%%%%%%%%%%%%%%%%%%%%%%%%%%%%%%%%%%%

\end{document}